\documentclass[aps,prl,twocolumn]{revtex4}
\usepackage{amssymb}
\usepackage{amsfonts}
\usepackage{amsmath}
\usepackage{graphicx}
\begin{document}


\title{Magnetotransport properties of the type II Weyl semimetal candidate Ta$_3$S$_2$}

\author{D. Chen$^{1,2}$, L. X. Zhao$^2$, J. B. He$^2$, H. Liang$^2$, S. Zhang$^2$, C. H. Li$^2$, L. Shan$^{2,3}$, S. C. Wang$^{1}$, Z. A. Ren$^{2,3}$, C. Ren$^{2,5}$, and G. F.
Chen$^{2,3,4\dag}$}

\affiliation{$^1$Department of Physics, Renmin University of China,
Beijing 100872, China}

\affiliation{$^2$Institute of Physics and Beijing National
Laboratory for Condensed Matter Physics, Chinese Academy of
Sciences, Beijing 100190, China}

\affiliation{$^3$Collaborative Innovation Center of Quantum Matter,
Beijing, China}

\affiliation{$^4$School of Physical Sciences, University of Chinese
Academy of Sciences, Beijing 100190, China}

\affiliation{$^5$Physics Department, Yunnan University, Kunming 671000, China}

\begin{abstract}
We have investigated the magnetoresistance (MR) and Hall resistivity
properties of the single crystals of tantalum sulfide, Ta$_3$S$_2$,
which was recently predicted to be a new type II Weyl
semimetal. Large MR (up to $\sim$
8000\% at 2 K and 16 T), field-induced metal-insulator-like
transition and nonlinear Hall resistivity are observed at low temperatures. The large MR shows a
strong dependence on the field orientation, leading to a giant
anisotropic magnetoresistance (AMR) effect. For the field applied
along the $b$-axis ($B\parallel b$), MR exhibits quadratic field
dependence at low fields and tends towards saturation at high
fields; while for $B\parallel a$, MR presents quadratic field
dependence at low fields and becomes linear at high fields without
any trend towards saturation. The analysis of the Hall resistivity
data indicates the coexistence of a large number of electrons with low mobility and a small number of holes with high mobility.
Shubnikov-de Haas (SdH) oscillation analysis reveals three
fundamental frequencies originated from the three-dimensional (3D) Fermi surface (FS)
pockets. We find that the semi-classical multiband model is sufficient to account for the experimentally observed MR in Ta$_3$S$_2$.
\end{abstract}

\maketitle

\newpage

\section{Introduction}
The recent discovery of Weyl and Dirac semimetals (WSM/DSM) has
attracted much attention in condensed matter physics and material
science \cite{a1,a2,a3,a4,a5,a6,a7,a8,a9,a10,a11}. These materials, having a
linear dispersion at the crossing of the valence and conduction
bands near the Fermi energy, can be considered as three-dimensional (3D)
analogues of graphene \cite{a12}. By breaking either time-reversal
symmetry or inversion symmetry, DSM can also evolve to WSM and the
Weyl nodes with specific chirality can be viewed as the magnetic
monopole in momentum space \cite{a13}. The WSM has been classified
into two types \cite{a7,a14}. The type I WSM exhibits ideal
conical Weyl cone with Lorentz symmetry. While in type II WSM, the
Weyl cone is extreme tilted and Lorentz symmetry is violated.
Experimentally, Fermi arc has been detected by the angle resolved
photoemission spectroscopy (ARPES) measurements for both two types
of materials \cite{a9,a10,a15,a16}, while the negative longitudinal magnetoresistance
(MR) has been observed only for the type I WSM \cite{a8,a17,a18}. Although some exotic
properties like novel anomalous Hall effect (AHE) has been
theoretically predicted for the type II WSM \cite{a19}, it is
not experimentally verified yet. It remains an important challenge to find
more new materials and to study the unusual magnetotransport
properties and the unique electronic structure.

\begin{figure}
\includegraphics[width=8cm, height=9.6cm]{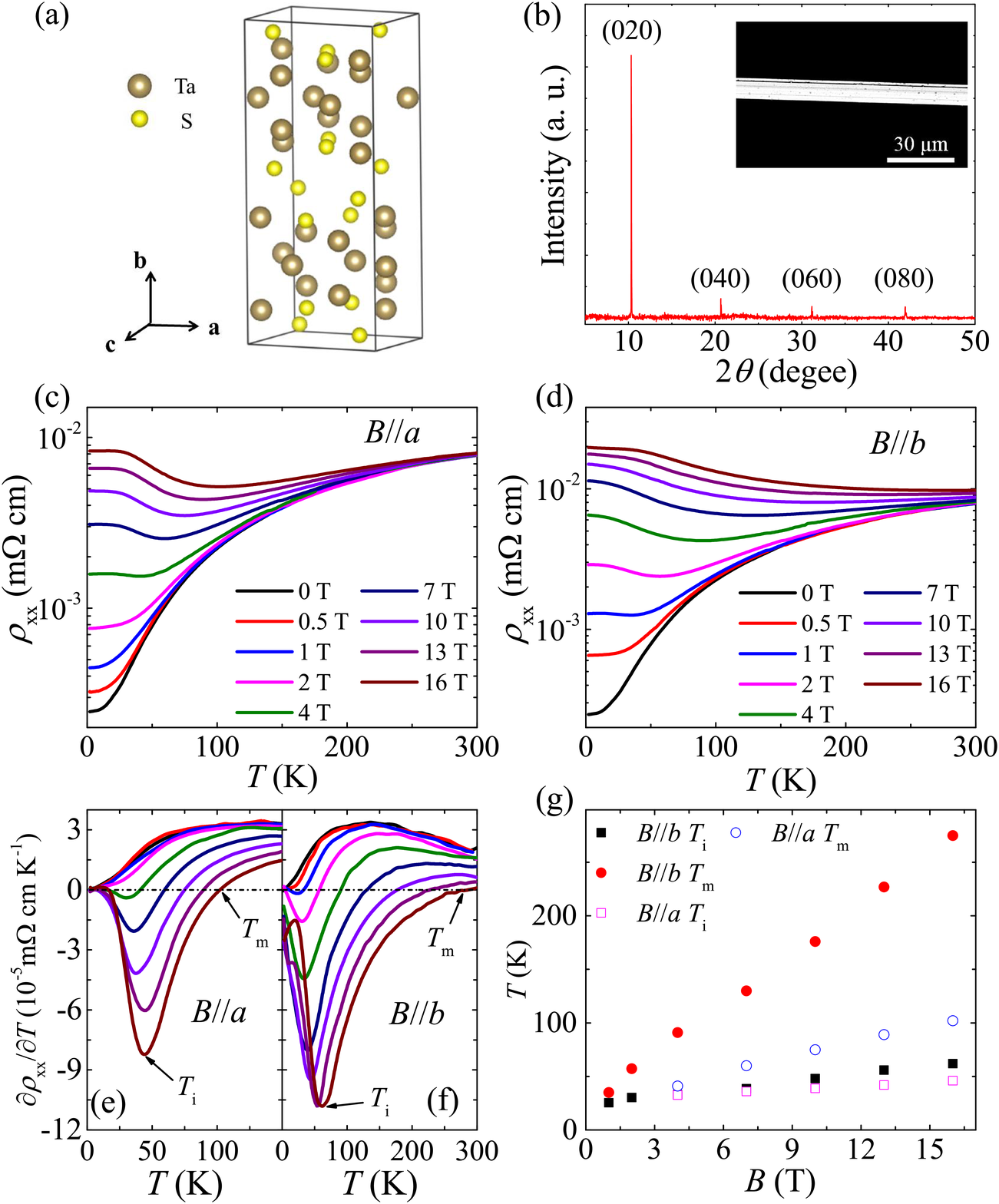}
\caption{\label{fig:fig1}(Color online)(a) The crystal structure of Ta$_3$S$_2$ with $Abm2$ space group.  The brown-gray and yellow balls represent the Ta and S atoms, respectively. (b) The XRD pattern of the Ta$_3$S$_2$ single crystal with $(0k0)$ reflections. Inset: A SEM image of a ribbon-shaped sample. (c)-(d) Temperature dependence of $\rho_{xx}$ under selected magnetic fields plotted on semi-log scale for $B\parallel a$ and $B\parallel b$, respectively. (e)-(f) The derivative of resistivity $\partial\rho_{xx}/\partial T$ for $B\parallel a$ and $B\parallel b$, respectively.  $T_m$ and $T_i$ are defined as the temperatures at which the $\partial \rho_{xx}/\partial T$ vs. $T$ curves change sign and take minimum, respectively.  (g) Magnetic field dependence of $T_m$ and $T_i$.}
\end{figure}

Tantalum sulfide Ta$_3$S$_2$ was predicted to be a robust type II
WSM candidate with 8 Weyl nodes \cite{a20}. The $k$-space separation
between Weyl nodes of Ta$_3$S$_2$ is the largest among the known
Weyl semimetal candidates. As shown in Fig. 1(a), Ta$_3$S$_2$
crystalizes in an orthorhombic structure without inversion symmetry,
which is the key to realizing the Weyl semimetal state. Transport
properties of polycrystalline Ta$_3$S$_2$ have been reported a few
decades ago, which show a semimetal behavior with MR up to about
100\% at low temperatures \cite{a21}. In this work, we have
successfully grown the single crystals of Ta$_3$S$_2$ and performed the
magnetotransport studies. We found a large positive MR = $[\rho_{xx}(B)-\rho_{xx}(0)]/\rho_{xx}(0)\times 100$\% up to 8000\%,
which is about two orders of magnitude higher than that of polycrystalline samples. The MR shows a strong anisotropic behavior when the
magnetic field is rotated in the $ab$-plane. Hall resistivity
measurement indicates that hole-type carriers with very high
mobility dominate the transport properties at low temperatures, and
a two-band model is sufficient to describe the large MR and its
deviation from the quadratic field dependence. The analysis of the Shubnikov-de Haas
(SdH) oscillations also reveals three fundamental frequencies
originated from the 3D Fermi surface (FS) pockets. All these results
reveal a complex electronic structure in Ta$_3$S$_2$.

\begin{figure}
\includegraphics[width=8cm, height=8cm]{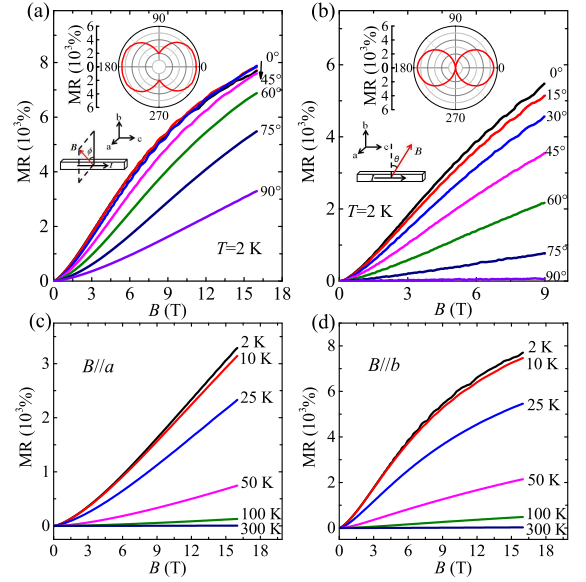}
\caption{\label{fig:fig1}(Color online)(a) Angular dependence of the MR with keeping magnetic field perpendicular to the electric current.  The lower inset depicts the measurement configuration, where $\phi$ is defined as the angle between the direction of the magnetic field and the $b$-axis of the sample.  The upper inset presents polar plot of the angular dependence of the MR as the magnetic field is rotated in $ab$ plane at 2 K and 9 T. (b) MR with the magnetic field applied from perpendicular ($\theta=0^{\circ}$) to parallel ($\theta=90^{\circ}$) to the electric current.  The lower inset depicts the measurement configuration, where $\theta$ is defined as the angle between the magnetic field and the $b$-axis.  The upper inset shows polar plot of the angular dependence of MR, where the magnitude is taken at 2 K and 9 T.  (c)-(d) Magnetic field dependence of MR at various temperatures for $B\parallel a$ and $B\parallel b$, respectively. }
\end{figure}

\section{Experiment and method}
The single crystals of Ta$_3$S$_2$ were grown by the chemical vapor
transport method \cite{a22}. Stoichiometric amounts of TaS$_2$ and
Ta powders were sealed in a Ta crucible with TaBr$_5$ as the transport
agent. The Ta crucible was then sealed in an evacuated quartz tube
and the quartz tube was heated and kept at 1000 $^{\circ}$C for 14 days before
turning the furnace off. The obtained crystals were characterized by
X-ray diffraction (XRD) on a PANalytical diffractometer with Cu
$K\alpha$ radiation at room temperature. MR and Hall resistivity measurements were performed using a Quantum Design physical property measurement system (PPMS).

The obtained ribbon-shaped crystals are $10\sim 15$ mm in length,
$5\sim 10$ $\mu$m in width, and $2\sim 5$ $\mu$m in thickness. Powder
XRD measurement confirmed that the obtained crystals
crystalize in an orthorhombic structure with $Abm2$ space group. The
average Ta:S atomic ratio determined by energy-dispersive X-ray (EDX) spectroscopy analysis is close to 3:2. Fig. 1(b) shows the XRD pattern of a single crystal sample. All of the peaks can be identified as the ($0k0$) reflections of Ta$_3$S$_2$. The inset of Fig. 1(b) shows the scanning electron microscopy (SEM) image of
Ta$_3$S$_2$ with $c$-axis being the growth direction of the ribbon.

\section{Result and discussion}
\begin{figure}
\includegraphics[width=8cm, height=8cm]{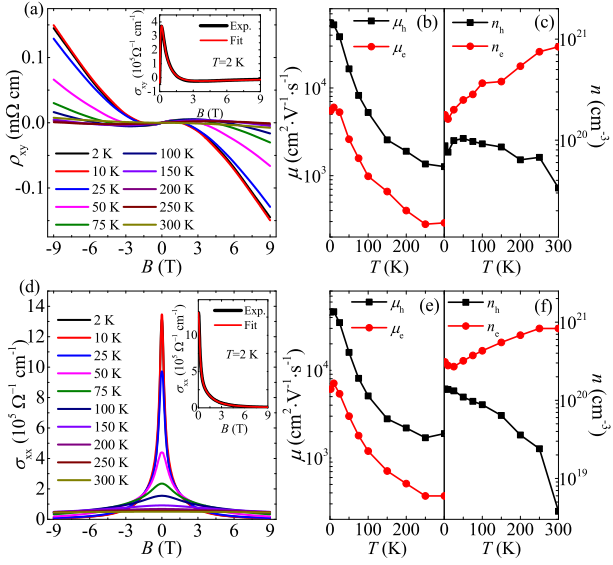}
\caption{\label{fig:fig1}(Color online)(a) The magnetic field dependence of Hall resistivity $\rho_{xy}$ at various temperatures. The inset shows the experimental data of Hall conductivity at 2 K and the fitted curve. (b)-(c) The temperature dependence of carrier densities and carrier mobilities of electrons and holes obtained by fitting Hall conductivity $\sigma_{xy}$. (d) The magnetic field dependence of longitudinal conductivity $\sigma_{xx}$. The inset shows the experimental data of $\sigma_{xx}$ at 2 K and the fitted curve. (e)-(f) The temperature dependence of carrier densities and carrier mobilities of electrons and holes obtained by fitting $\sigma_{xx}$.}
\end{figure}

We carried out the temperature dependence of the resistivity $\rho_{xx}$ measurements with magnetic field $B$ parallel to the crystallographic $a$-axis and to $b$-axis, as shown in Fig. 1(c) and (d), respectively. It is noted that in both configurations, $B$ is always perpendicular to electric current $I$.  At zero field, $\rho_{xx}$ shows a metallic behavior with a residual resistivity ratio (RRR) up to 30, which is much higher than that of polycrystalline sample \cite{a21}. With the application of magnetic field, $\rho_{xx}$ is enhanced drastically at low temperatures and shows a minimum, indicating that this system undergoes a field induced metal-insulator like transition. Remarkably, $\rho_{xx}$ saturates at very low temperatures, and leads to a resistivity plateau. The magnetic field induced metal-insulator like transition and resistivity plateaus has been observed in many semimetals, including the recent discovered compounds like WTe$_2$, TaAs, LaSb, PtSn$_4$ \cite{a8,a11,a23,a24} as well as the well-studied elements like graphite and bismuth \cite{a25,a26}, but its underlying mechanism is still under debate. The resistivity plateaus have also been observed in topological insulators (TIs) such as Bi$_2$Te$_2$Se and SmB$_6$ in which the resistivity plateau is considered to originate from the short-circuiting effect of the metallic surface state on the bulk insulating state \cite{a27,a28}. To get a more clear view of the exotic transport behavior, we plotted the derivative $\partial \rho_{xx}/\partial T$ curves at several magnetic fields in Fig. 1(e) and (f), respectively, where $T_m$ and $T_i$ are determined as the temperatures at which $\partial \rho_{xx}/\partial T$ versus $T$ curves change sign and take minimum, which correspond to the temperatures of the emergence of metal-insulator like transition and resistivity plateau in $\rho_{xx}$ curves. The magnetic field dependence of the $T_m$ and $T_i$ is shown in Fig. 1(g). For both $B\parallel a$ and $B\parallel b$, $T_m$ increases rapidly with increasing field, whereas $T_i$ is almost unchanged even at high fields. The tendency of the magnetic field induced metal-insulator like transition for $B\parallel b$ is larger than that for $B\parallel a$. The metal-insulator like transition was tentatively ascribed to multiband effects in the system with low carrier density and electron-hole compensation \cite{a29}, or an excitonic gap induced by magnetic field \cite{a30}. However, none of the existing theories seem to account for the saturation of the low temperature resistance in a fixed magnetic field in semimetals.

\begin{figure}
\includegraphics[width=9cm, height=7.5cm]{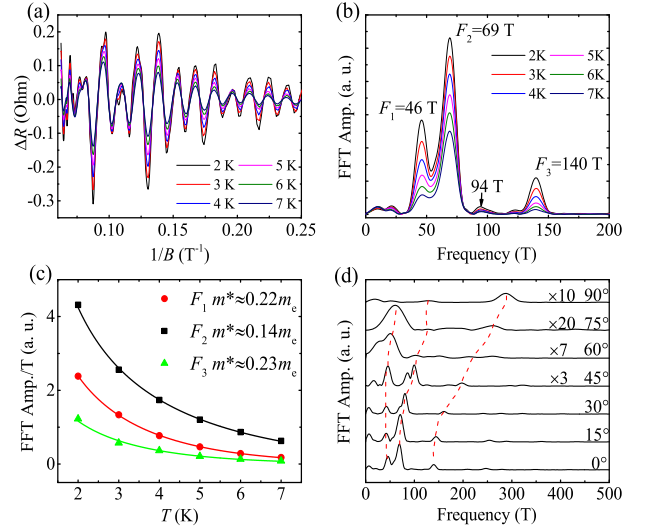}
\caption{\label{fig:fig1}(Color online)(a) SdH oscillations after subtracting background at 2-7 K for $B\parallel b$ as a function of $1/B$. (b) The FFT spectra of oscillations at 2-7 K with three fundamental frequencies. (c) The oscillation amplitudes as function of temperature with fitted curves. (d) Angular dependence of the FFT spectra of the oscillations taken at 2 K by rotating the field from parallel to perpendicular to $b$-axis. The dashed lines are guides to the eyes.}
\end{figure}

The upper inset of Fig. 2(a) presents polar plot of the angular dependence of the MR measured at 2 K and 9 T as the electric current flows along the crystallographic $c$-axis, and the magnetic field is rotated in $ab$-plane. The lower inset depicts the measurement configuration, where $\phi$ is defined as the angle between the direction of the magnetic field and the $b$-axis of the sample. This dipole-like pattern indicates that the MR for magnetic field along $b$-axis is much larger than that for magnetic field along $a$-axis. Fig. 2(a) shows the field dependence of MR at different $\phi$ up to 16 T. A large MR up to 8000\% is observed for $\phi=0^{\circ}$. Near the zero field, MR presents a quadratic field dependence before a linear behavior at moderate fields, after which MR shows a tendency to
saturation up to 16 T. The obvious SdH oscillations indicate the high quality of the sample. One can see that the SdH oscillations are suppressed remarkably as $\phi$ is increased. The field dependence of the MR is also measured at 2 K by tilting the magnetic field from perpendicular to parallel to the applied current, as shown in Fig. 2(b). Dramatic drop of the MR is observed but no tendency of negative MR is detected up to 9 T for $B\parallel I$. The angular dependence can be more directly seen in the upper inset of Fig. 2(b), where the magnitude of the MR is taken at 9 T and 2 K.

Figure 2(c) and (d) show the magnetic field dependence of MR at various temperatures for $B\parallel a$ and $B\parallel b$ with keeping the magnetic field perpendicular to the electric current, respectively. In contrast to the high field saturation of low temperature MR for the magnetic field applied along $b$-axis ($B\parallel b$), MR
exhibits quadratic field dependence at low fields and becomes linear at high fields without any trend towards saturation for the field applied along $a$-axis ($B\parallel a$). With increasing temperature, MR drops off obviously for both two directions. Below 10 K, obvious SdH oscillations are present in the MR curves for $B\parallel b$
above 4 T, while the SdH oscillations for $B\parallel a$ are very weak and present only above 10 T. The difference of the amplitude of the SdH oscillations between the two directions indicates that the difference of the carrier mobility and effective mass between the two directions.

To provide some clues to understand the exotic magnetotransport properties, we performed Hall-effect measurement on the single crystal of Ta$_3$S$_2$. The magnetic field is approximately along $b$-axis. Figure 3(a) and (d) present the magnetic field dependence of Hall resistivity $\rho_{xy}$ and longitudinal conductivity $\sigma_{xx}$ at several temperatures from 2 K to 300 K. At low temperatures, $\rho_{xy}$ has a positive slope in low field and a negative slope
in high field, suggesting the coexistence of two types of carriers. At higher temperatures, the positive slope composition gradually decays and the $\rho_{xy}$ curves tend to be linear, suggesting that electrons dominate the main transport processes. All these are consistent with multiple hole- and electron-like carriers as observed in other semimetals \cite{a6,a8,a31}. In a semiclassical two-band model, the Hall conductivity
$\sigma_{xy}$ and longitudinal conductivity $\sigma_{xx}$ are expressed
as \cite{a32,a33}:
\begin{equation}
\sigma_{xy}=[\frac{n_h \mu_h^2}{1+(\mu_hB)^2} -
\frac{n_e\mu_e^2}{1+(\mu_eB)^2}]eB,
\end{equation}
\begin{equation}
\sigma_{xx}=\frac{e n_e \mu_e}{1+(\mu_e B)^2} + \frac{\sigma_{xx}(0) - e
n_e \mu_e}{1+(\mu_h B)^2},
\end{equation}
Where $n_e$ (or $n_h$) is the carrier density of electron (or hole), $\mu_e$ (or $\mu_h$) is the mobility of electron (or hole), respectively. We find that, both of $\sigma_{xy}$ and $\sigma_{xx}$ can be well fitted by the two-band model. The fitting curves for $T$=2 K are shown in the inset of Fig. 3(a) and (d), respectively. The carrier densities and mobilities obtained by fitting $\sigma_{xy}$ and $\sigma_{xx}$ have same order of magnitude and
similar temperature dependent tendency, as shown in Fig. 3(b), (c), (e), and (f). A good agreement between the fitting of $\sigma_{xy}$ and $\sigma_{xx}$ indicates that two-band model is sufficient to describe the behavior of the field dependent MR. Both of the mobilities of the two type carriers decrease with increasing temperature. At $T$=2 K, the mobility of holes, $\mu_h \simeq 5 \times 10^4$ cm$^2$ V$^{-1}$s$^{-1}$, is larger than that of electrons $\mu_e$ by one order of magnitude. Furthermore, as the temperature decreases, the electron density decreases and the hole density increases. The two carrier densities tend to be close to each other but have no possibility to compensate even at extreme low temperature. Remarkably, our results suggest that perfect
electron-hole compensation is not be necessary for the large MR in Ta$_3$S$_2$, and classical theory of magnetoresistance is sufficient to describe the quadratic field dependence of MR at low fields and
saturation of MR at high fields as seen in normal metals \cite{a34}.

SdH oscillations are proved to be a powerful tool in the studies of topological quantum materials. Figure 4(a) shows the plots of oscillatory components $\Delta R$ versus $1/B$ at several temperatures, obtained by subtracting a background from the $R(B)$ isotherms. The fast Fourier transform (FFT) analysis of these data yields three fundamental frequencies at about $F_1$ = 46 T, $F_2$ = 69 T, and $F_3$ = 140 T, and the weak peak at 94 T is
the second harmonics of $F_1$, as shown in Fig. 4(b). The cyclotron masses for the three frequencies can be obtained by fitting the temperature dependent FFT amplitude using the Lifshitz-Kosevitch formula \cite{a35}: $ A \propto \frac{2\pi^2k_BTm^{*}/eB\hbar}{\sinh(2\pi^2k_BTm^{*}/eB\hbar)}$, where $k_B$ is the Boltzmann constant and $m^{*}$ is cyclotron mass. $B$ is determined by the range of oscillation:
$1/B=(1/B_1+1/B_2)/2$, where $1/B_1$ and $1/B_2$ are the up and down limits of $1/B$. The fitting result is shown in Fig. 4(c). The cyclotron masses for the three frequencies are estimated to be 0.22$m_e$, 0.14$m_e$, and 0.23$m_e$, respectively. The cross-sectional area of the FS can be obtained by the Onsager relation \cite{a35}: $F=(\Phi_0/2\pi^2)A_F$, where $F$ is the FFT frequency, $A_F$ is the cross-sectional area of the FS and $\Phi_0$
is the flux quantum. The cross-sectional areas of the three frequencies are $4.4\times 10^{-3}$ {\AA}$^{-2}$, $6.6\times 10^{-3}$ {\AA}$^{-2}$, and $1.3\times 10^{-2}$ {\AA}$^{-2}$, occupying 0.47\%,
0.7\%, and 1.4\% of the cross-sectional area of the first Brillouin zone (BZ), respectively. The small ratio of the cross-sectional area of FS in the first BZ is the feature of semimetals. Fig. 4(d) shows the FFT spectra of the SdH oscillations measured at 2 K by rotating the field from parallel to perpendicular to $b$-axis. We note that,
$F_1$ is almost unchanged, indicating a spherical FS pocket. The other two frequencies change monotonically with rotating the magnetic field, indicating the ellipsoidal pockets.

\section{Conclusion}
In conclusion, we have successfully grown the single crystals of
Ta$_3$S$_2$, a robust type II WSM candidate. Large MR and
anisotropic MR are observed when the magnetic field is applied
perpendicular to current. The temperature dependent resistivity shows the magnetic
field induced metal-insulator like transition and resistivity
plateau at low temperature. The metal-insulator like transition
temperature increases rapidly with increasing field, whereas the
temperature of saturation of the resistivity is almost independent
with magnetic field. Hall resistivity measurement reveals the high
mobility of holes without the compensation of electrons and holes.
The analysis of SdH oscillations reveals
three FS pockets including a sphere pocket and two ellipsoidal
pockets. We hope this work can provide valuable clues for further research on
Ta$_3$S$_2$ and other type II Weyl semimetals.

\textbf{Acknowledgements:} This work was supported by National Basic
Research Program of China 973 Program (Grant No. 2015CB921303), and
the ``Strategic Priority Research Program (B)'' of the Chinese
Academy of Sciences (Grant No. XDB07020100).

$\dag$gfchen@iphy.ac.cn


\begin{thebibliography}{99}

\bibitem{a1}X. Wan, A. M. Turner, A. Vishwanath, and S. Y. Savrasov, Phys. Rev. B \textbf{83}, 205101
(2011).

\bibitem{a2}A. A. Burkov and L. Balents, Phys. Rev. Lett. \textbf{107}, 127205
(2011).

\bibitem{a3}Z. Wang, Y. Sun, X. Q. Chen, C. Franchini, G. Xu, H. Weng, X. Dai, and Z. Fang, Phys. Rev. B \textbf{85}, 195320
(2012).

\bibitem{a4}Z. K. Liu, B. Zhou, Y. Zhang, Z. J. Wang, H. M. Weng, D. Prabhakaran, S. K. Mo, Z. X. Shen, Z. Fang, X. Dai, Z. Hussain, and Y. L. Chen, Science \textbf{343}, 864
(2014).

\bibitem{a5}Z. Wang, H. Weng, Q. Wu, X. Dai, and Z. Fang, Phys. Rev. B \textbf{88}, 125427
(2013).

\bibitem{a6}T. Liang, Q. Gibson, M. N. Ali, M. Liu, R. J. Cava, and N. P. Ong, Nat. Mater. \textbf{14}, 280
(2015).

\bibitem{a7}H. Weng, C. Fang, Z. Fang, B. A. Bernevig, and X. Dai, Phys. Rev. X \textbf{5}, 011029
(2015).

\bibitem{a8}X. Huang, L. Zhao, Y. Long, P. Wang, D. Chen, Z. Yang, H. Liang, M. Xue, H. Weng, Z. Fang, X. Dai, and G. Chen, Phys. Rev. X \textbf{5}, 031023
(2015).

\bibitem{a9}S. Y. Xu, I. Belopolski, N. Alidoust, M. Neupane, G. Bian, C. L. Zhang, R. Sankar, G. Q. Chang, Z. J. Yuan, C. C. Lee, S. M. Huang, H. Zheng, J. Ma, D. S. Sanchez,
B. K. Wang, A. Bansil, F. C. Chou, P. P. Shibayev, H. Lin, S. Jia,
and M. Z. Hasan, Science \textbf{349}, 613 (2015).

\bibitem{a10}B. Q. Lv, H. M. Weng, B. B. Fu, X. P. Wang, H. Miao, J. Ma, P. Richard, X. C. Huang, L. X. Zhao, G. F. Chen, Z. Fang, X. Dai, T. Qian, and H. Ding, Phys. Rev. X \textbf{5}, 031013
(2015).

\bibitem{a11}M. N. Ali, J. Xiong, S. Flynn, J. Tao, Q. D. Gibson, L. M. Schoop, T. Liang, N. Haldolaarachchige, M. Hirschberger, N. P. Ong, and R. J. Cava, Nature \textbf{514}, 205
(2014).

\bibitem{a12}K. S. Novoselov, A. K. Geim, S. V. Morozov, D. Jiang, M. I. Katsnelson, I. V. Grigorieva, S. V. Dubonos, and A. A. Firsov, Nature \textbf{438}, 197
(2005).

\bibitem{a13}Z. Fang, N. Nagaosa, K. S. Takahashi, A. Asamitsu, R. Mathieu, T. Ogasawara, H. Yamada, M. Kawasaki, Y. Tokura, and K. Terakura, Science \textbf{302}, 92
(2003).

\bibitem{a14}A. A. Soluyanov, D. Gresch, Z. Wang, Q. Wu, M. Troyer, X. Dai, and B. A. Bernevig, Nature \textbf{527}, 495
(2015).

\bibitem{a15}K. Deng, G. Wan, P. Deng, K. Zhang, S. Ding, E. Wang, M. Yan, H. Huang, H. Zhang, Z. Xu, J. Denlinger, A. Fedorov, H. Yang, W. Duan, H. Yao, Y. Wu, S. Fan, H. Zhang, X. Chen, and S. Zhou, arXiv:1603.08508.

\bibitem{a16}Y. Wu, N. H. Jo, D. Mou, L. Huang, S. L. Bud'ko, P. C. Canfield, and A. Kaminski, arXiv:1604.05176.

\bibitem{a17}F.  Arnold, C. Shekhar, S. Wu, Y. Sun,  R. D. d. Reis, N. Kumar, M. Naumann, M. O. Ajeesh, M. Schmidt, A. G. Grushin, J. H. Bardarson, M. Baenitz, D. Sokolov, H. Borrmann, M. Nicklas, C. Felser, E. Hassinger, and B. Yan, Nat. Comms. \textbf{7}, 11615 (2016)

\bibitem{a18}X. Yang, Y. Li, Z. Wang, Y. Zhen, and Z. Xu, arXiv:1506.02283.

\bibitem{a19}  A. A. Zyuzin and R. P. Tiwari, arXiv:1601.00890.

\bibitem{a20}G. Chang, S. Y. Xu, D. S. Sanchez, S. M. Huang, C. C. Lee, T. R. Chang, G. Bian, H. Zheng, I. Belopolski, N. Alidoust, H. T. Jeng, A. Bansil, H. Lin, and M. Z. Hasan, Sci. Adv. \textbf{2}, e1600295
(2016).

\bibitem{a21}H. Nozaki, H. Wada, and S. Takekawa, J. Phys. Soc. Jpn. \textbf{60}, 3510
(1991).

\bibitem{a22}S. J. Kim, K. S. Nanjundaswamy, and T. Hughbanks, Inorg. Chem. \textbf{30}, 159
(1991).

\bibitem{a23} F. F. Tafti, Q. D. Gibson, S. K. Kushwaha, N. Haldolaarachchige, and R. J. Cava,
Nat. Phys. \textbf{12}, 272 (2015).

\bibitem{a24}E. Mun, H. Ko, G. J. Miller, G. D. Samolyuk, S. L. Bud'ko, and P. C. Canfield, Phys. Rev. B \textbf{85}, 035135
(2012).

\bibitem{a25}D. E. Soule, Phys. Rev. \textbf{112}, 698 (1958).

\bibitem{a26}Z.  Zhu,  A.  Collaudin,  B.  Fauqu¨¦,  W.  Kang,  and  K.  Behnia,  Nat.  Phys. \textbf{8}, 89
(2011).

\bibitem{a27} D. J. Kim, S. Thomas, T. Grant, J. Botimer, Z. Fisk, and J. Xia, Sci. Rep. \textbf{3}, 3150
(2013).

\bibitem{a28}Z. Ren, A. A. Taskin, S. Sasaki, K. Segawa, and Y. Ando, Phys. Rev. B \textbf{82}, 241306(R)
(2010).

\bibitem{a29}X. Du, S. W. Tsai, D. L. Maslov, and A. F. Hebard, Phys. Rev. Lett. \textbf{94}, 166601
(2005).

\bibitem{a30}D. V. Khveshchenko, Phys. Rev. Lett. \textbf{87}, 206401
(2001).

\bibitem{a31}C. Shekhar, A. K. Nayak, Y. Sun, M. Schmidt, M. Nicklas, I. Leermakers, U. Zeitler, Y. Skourski, J. Wosnitza, Z. Liu, Y. Chen, W. Schnelle, H. Borrmann, Y. Grin, C. Felser, and B. Yan, Nat. Phys. \textbf{11}, 645
(2015).

\bibitem{a32}N. W. Ashcraft and N. D. Mermin, \emph{Solid Sate Physics} (Holt, Rinehart and Winston, New York,
1976).

\bibitem{a33}C. M. Hurd, \emph{The Hall Effect in Metals and Alloys} (Cambridge University Press,
Cambridge, 1972).

\bibitem{a34}A. A. Abrikosov, J. Phys. A: Math. Gen. \textbf{36}, 9119 (2003).

\bibitem{a35}D. Shoenberg, \emph{Magnetic Oscillations in Metals} (Cambridge University Press, Cambridge,
1984).

\end{thebibliography}
\end{document}